\documentclass[a4paper, prb, reprint, hyphens, citeautoscript]{revtex4-1}
\usepackage{graphicx}
\usepackage{geometry}

\makeatletter
\def\l@subsubsection#1#2{}
\makeatother

\begin{document}

\title{Transforming physics laboratory work from `cookbook' type to genuine inquiry}
\date{\today}
\author{K. Dunnett}
\email{kirsty.dunnett@su.se}
\author{M. K. Kristiansson}
\author{G. Eklund}
\author{H. {\"O}str{\"o}m}
\author{A. Rydh}
\author{F. Hellberg}
\affiliation{Department of Physics, Stockholm University, SE-106 91 Stockholm, Sweden}

\begin{abstract}
`Cookbook' style laboratory tasks have long been criticised for the lack of critical and independent thought that students need in order to complete them. We present an account of how we transformed a 'cookbook' lab to a genuine inquiry experiment in first year physics. Crucial features of the work were visits to see other teaching laboratories, understanding student preparedness and the selection of an appropriate experiment to develop. The new two session laboratory work is structured so students make decisions related to the method of a basic experiment in the first session and then have freedom to investigate any aspect they wish to in the second. Formative feedback on laboratory notebook keeping is provided by short online activities.
\end{abstract}

\maketitle

\tableofcontents

\section{Introduction}

A common underlying assumption of practical work in physics courses is that it provides students with opportunities to develop practical skills (e.g. equipment use) and reinforces knowledge by allowing students to interact directly with a physical realisation of theoretical concepts \cite{SciEd.88.28, RevEdRes.52.201}. Laboratory work with the express purpose of reinforcing lecture content often has very precise instructions, with very little scope for students to deviate, and the aim is usually to obtain a known result \cite{AJP.47.859, JColSciTeach.38.52}. These `cookbook' style laboratory tasks have long been criticised for the lack of critical and independent thought that students need in order to complete them \cite{SciEd.88.28, AJP.47.859, ThePhysTeach.53.6.349}. 

More recently, the effect of different styles of laboratory work on the development of expert-like beliefs has been studied in some detail, with the conclusion that laboratory courses that allow for some element of independent inquiry promote the development of expert-like beliefs \cite{PRPER.12.020132}. On the other hand, laboratory work with the sole purpose of demonstrating a particular phenomenon described in lectures is not effective \cite{PRPER.13.010129}, but may actually lead to \textit{less} expert beliefs about the nature of experimentation \cite{PRPER.12.020132}. In addition, more open-ended laboratories may reduce gender differences in participation \cite{JColSciTeach.31.225}.

To our knowledge, creating an inquiry laboratory from more or less `cookbook' experiments is a largely undocumented process, with the resulting new laboratory course or implementation typically being presented \cite{EJP.39.025702, EJP.39.025703, Scient-educ.8, PhysTeach.55.159}, sometimes with a description of the situation that prompted the changes \cite{JColSciTeach.40.45}. Here, we provide a record of how we transformed the two laboratory sessions associated with an introductory thermodynamics course from prescribed activities (`cookbook' experiments) to a genuine experiment (open-ended inquiry). We discuss our decision making process in some detail and the motivation for each aspect of the new laboratory work at Stockholm University (SU).

\section{Preparing for the changes}
Before making any changes to the laboratory work, it was important to gain a broad perspective on laboratory work to ensure a successful outcome. As well as reports in the literature, and developing our understanding of our own students, visits to teaching laboratories at other universities, to see other ways of arranging and delivering experimental work, provided much inspiration of what might be possible.

\subsection{Understanding the students}

A survey of first year students at the beginning of their laboratory work in autumn 2018 found that the majority already had some experience of making decisions about doing experimental work (as summarised in Table \ref{tab:PriorExp}). As such, they were considerably better prepared for independent experimental work than many of the existing laboratory tasks assumed. This was consistent with the positive feedback on the one laboratory for which they made significant decisions on the experimental process \cite{Optics-2019-review}. While students may not have experience of using equipment such as oscilloscopes, the existing laboratory work with its prescribed methods could be restricting students' development.

\begin{table}[tb!]
\begin{tabular}{p{0.6\columnwidth}ccc}
\textbf{Activity/Prior experience} & \textbf{Lots} & \textbf{Some} & \textbf{None} \\ \hline
Designed, built and conducted own experiments & 12 & 25 & 3 \\ \hline
Conducted set experiments with own method & 6 & 22 & 12 \\ \hline
Conducted set experiments with prescribed method & 24 & 15 & 0 \\ \hline
Took data while teacher demonstrated experiments & 7 & 20 & 13 \\ \hline
Analysed data from an experiment I did not conduct & 3 & 25 & 12 \\ \hline
Experience of research labs \newline (industrial/academic) & 0 & 5 & 35 \\ \hline
\end{tabular}
\caption{Summary of SU students' prior experience with laboratory work, autumn 2018: the majority of students had some experience of making decisions when doing physics experiments. \label{tab:PriorExp} }
\end{table}

\subsection{Existing situation}

In 2013, the experimental part of Stockholm University's Bachelor program in physics was revised along with the rest of the program. Before then, the first year had included a course that covered data handling, statistics, error analysis, and report writing. Students found this course difficult, and many did not pass on the first attempt. When the program was restructured, a progressive development of experimental work from straight-forward activities in first year to independent project work in the third year was introduced. 

The first year of the physics program now consists of a five-part course on Classical Physics. Each part, except for the first one, includes one or two laboratory sessions. In each laboratory session, students perform experiments connected to the theory in the lectures (a `labs in course' arrangement). The intention is to motivate students with interesting experiments and for them to learn some basic experimental skills. Statistics, data analysis (including uncertainties) and report writing are now covered in the second year.

In 2015-16, the experimental part of the physics program was reviewed, with the conclusion that further improvements could be beneficial to the students. The laboratory work associated with the first year was identified as needing particular work, and that the thermodynamics laboratory should be prioritised.

In 2018 a working group was formed to improve the laboratory work from the results of that study. For example, the purpose of the labs in the first year course Classical Physics is not explicitly specified in the course plan, and the value of classic `cookbook style' laboratory work, as found in many of the first year experiments, can be questioned \cite{SciEd.88.28, AJP.47.859,ThePhysTeach.53.6.349}. The working group consisted of the Deputy director of studies who was also responsible for the second year laboratory work (FH); the teaching laboratories coordinator (AR); the lecturer for the thermodynamics course (A\"O); the two post-graduate teaching assistants for the thermodynamics course (GE and MK); an experienced laboratory teaching assistant and developer (KD).

\subsection{Visits to other teaching laboratories}

Early in the discussion (autumn 2018), three of us (FH, AR, KD) visited the teaching laboratories at University College London (UCL) and Imperial College London (Imperial). We received guided tours of the different teaching laboratories and overviews of the aims and course structures along with various known issues and how they had been addressed. In addition, we met with many staff associated with the laboratories in different roles and had plenty of opportunity to discuss their thoughts about what had been done, and what might still be developed. The arrangement at UCL, was already familiar to one of us (KD had been a teaching assistant there for two years), while the laboratories at Imperial were new to all.

Beyond the examples of different teaching laboratories, these visits were particularly valuable for two reasons. 1) They provided exposure to development work that had been tried, tested and refined, but which hadn't always been documented or made public by publication. 2) We were discussing the process of developing laboratory work with those who had recently made significant changes (and investment) in their laboratory work (particularly at Imperial). 

The specific influences of these and other visits on the transformed laboratory work will be returned to later, for now, two points only will be noted, that are typical of UK Physics degree schemes. 1) The dedicated laboratory courses from the start of a degree programme contrasted with the `labs in course' arrangement at SU and typical in Sweden. 2) Laboratory notebook keeping was considered as a key skill for students to master, especially in the first year, but formed no part of the existing training at SU.

\section{Developing the inquiry laboratory}

With inspiration from the visits described above, it was decided to start with improving the labs in Thermodynamics course. It is the first laboratory work the students have at university, and had also been identified as most needing improvement.    

\subsection{Previous thermodynamics laboratory}

There were four existing experiments that we could choose between, the titles of which should indicate the topic; we include some notes on their relevance or interest later. All laboratory activities were highly structured, lying close to the `Confirmation' level of inquiry in which the entire process from problem to expected result is known before starting the experimental work, a fairly common situation for physics laboratories \cite{AJP.47.859,JColSciTeach.38.52}.

In the first laboratory session, students took measurements to determine either the specific heat capacity or thermal conductivity of a sample. Both experiments used the same equipment and sample materials. For the specific heat capacity experiment the samples were small cuboid blocks while for the thermal conductivity experiment the samples were rods about 15cm long. The latter experiment was less directly related to the content of the thermodynamics course than the former. 

In the second experimental session, students worked either with a heat pump or a gas thermometer. The heat pump `experiment' was known to be a bit dull, and a gas thermometer was considered more exciting, and therefore potentially more interesting for students. Unlike the experiments of the first laboratory session, these two experiments had considerably different equipment requirements.

Before each laboratory session, students completed preparatory activities for both experiments, although they would only complete one. These typically consisted of questions related to the theoretical background of the experiments, the workings of pieces of equipment, material properties (standard values), and sometimes also derivations of the formulae required to determine the quantity of interest from the measurements to be taken. In general, they prepared students for `what to expect' from the experimental activity. 

In the three hour laboratory sessions, the first half an hour or more was spent reviewing the preparatory activities, including the derivations, which often caused problems; students were thus provided with the correct answers in the laboratory session. The experiments were already set up and all students had to do was follow the instructions in the manual to obtain data points that were entered into spaces. They then performed the analysis steps as prompted. The demands of the experimental work were such as to ensure that students did not run out of time during the sessions. While all students could therefore complete the experiment, it meant that some could leave up to an hour early, and the challenges students faced were minimal, reducing success to meeting expectations, obtaining `good' values and completing previously defined tasks.

\subsection{Choice of experiment}

When we started the development work, we knew we wanted to develop some sort of inquiry laboratory, and provide some structured skill training, but had not made any firm decisions about the exact purpose, scope, format or structure of the new laboratory work. Various ideas were discussed in our initial meeting: Might all students either complete two experiments (one per session), or might half the students do one experiment and the others do the other one (different two session experiments), with some sort or presentation afterwards? It was first decided to focus on the specific heat capacity (with its relation to the lecture material) and gas thermometer experiments (with its fun aspects).

In this initial meeting, we also decided that it would be nice to introduce formal laboratory notebook keeping as part of the laboratory skills training, an important skill for active scientists \cite{PRPER.12.020129}. An additional potential advantage for students would be that this became their evidence not only of \textit{doing} experimental work, but also of \textit{recording} it systematically. However, at this stage, this was only a tentative decision, and it did not become clear that this would be realised until considerably later.

Further discussions a few days later revealed that the gas thermometer experiment did not permit a lot of avenues of investigation within the limits of the laboratory set up: the number of data points students would be able to obtain would not be sufficient for detailed or truly scientific investigation, or the appearance of the most interesting features. Between the two meetings, the two PhD students who would run the transformed laboratory sessions (MK and GE) performed the specific heat capacity experiment to test its suitability as an inquiry experiment. This experiment was not only the one most relevant to the lecture material, but also the most promising for transformation to an inquiry experiment as the data obtained showed a number of nuances or `Secret Objectives' \cite{ArXiv:1905.07267} that the students could find and investigate. While the `Secret Objectives' concept is not used in delivering the new experiment, it has been a useful conceptual tool and provides a convenient notation for staff discussions.

We had therefore reached the conclusion that students would perform one experiment, with the provisional aim to obtain the specific heat capacity of one or more samples, over two sessions. With the two sessions, students would be able to develop their experimental procedure, study different samples, or investigate their data in more depth. The question: `What would you do differently if you were to repeat the experiment?' was transformed from a hypothetical consideration into a serious proposition. Moreover, in keeping with any research experiment, the analysis of the experimental data became an activity that could be completed outside of the laboratory time. The new experiment was to prioritise using the scheduled laboratory time for data collection, and the rest of the development work would need to take this, and the students' workload from lectures, into account.

\subsection{Development concerns}

Having decided that a single experiment would be performed over two sessions separated by several weeks (determined by the class schedule), there were several structural issues that then needed to be addressed. Firstly, the theoretical derivation of the expression for the specific heat capacity, $c_p$, that had previously formed part of the preparatory activities for the session, was demanding. The laboratory manual stated that the preparatory activities for the two experiments (specific heat capacity and thermal conductivity) should take no more than two hours, but the derivation part alone took several hours, even for a researcher. The time constraints of the scheduled laboratory sessions (even when increased from three to four hours) introduced a priority to maximise the time students could spend collecting data, best achieved by minimising the amount of writing students had to do, especially in the form of copying or summarising introductory material from other laboratory documentation. 

In addition, a crucial question was how to ensure that students felt that they had succeeded during each stage of the laboratory work. This factor of success, coupled with knowledge that students struggled with the derivation that had previously formed part of the preparatory tasks led to the decision to provide the derivation for the students as part of the material provided. Despite the value of the authentic nature of inquiry laboratories both from students' perception \cite{IntJScholTandL.3.2.136} and for developing expert-like beliefs \cite{PRPER.12.020132}, students can find inquiry activities frustrating \cite{IntJScholTandL.3.2.136}. The new laboratories would need to be structured in such a way as to allow students freedom of investigation while ensuring that they were not overwhelmed by the need to make decisions. 

This brought the focus of the discussion onto the purpose of the laboratory work. We already had some idea of the direction our efforts might go in, but had yet to make any concrete decisions. In particular: beyond experience of genuine experimentation, what specific elements of experimentation did we want to introduce students to, and what skills did we want them to develop?
 
Two dominant features of laboratory work at UCL and Imperial were the introduction to laboratory notebook keeping, and learning to work with uncertainties and their propagation. This second topic is not taught until the second year at Stockholm University. Therefore, we did not want to emphasise uncertainties, although they form a crucial part of experimentation. The decision was made to ask for uncertainties as part of the raw data and provide the relevant uncertainty propagation formulae.

In the end, it was decided that the new laboratory work should provide a structured introduction to genuine experimental work, help students start developing experimental record keeping skills and promote habits of critically considering their data and results. The laboratories and their supporting activities were all designed to emphasise trying something, or looking to improve, avoiding pressure to meet set goals, and minimising the fear of failing by eliminating failure associated with an experiment not going to plan.

One last factor remains to be mentioned. The new laboratory work was to be self-contained, independent of the lecture course that provided the context of the experiment. Working with the course lecturer (H\"O) ensured that the aims were shared by those most concerned with the course delivery and facilitated communication with the students. Thus the new experiment does not `aim to allow students to improve their understanding of the specific heat capacity' by experience, but to \textit{experience performing an experiment within a thermodynamics context}. 

\subsection{Contextual developments}

Although the transformation of the thermodynamics experiment was performed as a stand-alone endeavour, the effort would result in a thermodynamics laboratory very different to the subsequent laboratory work in the first year, and the students deserved and would benefit from some explanation. In developing and defining the structure and purpose of the thermodynamics laboratory work that was our immediate concern, we also reviewed the overall structure and aims of the laboratory work throughout the entire degree programme, and understood how the laboratory work fitted together.

Thus, before creating the documentation immediately related to the new thermodynamics laboratory, we wrote more general documentation to provide the students with a more structured introduction and background to laboratory work. This ensured that the entire teaching team had a common, agreed understanding of laboratory work as it would be presented to the students, easing discussions of developing the experimental work. 

The parts of the documentation provided to first years were, in the order included in their laboratory pack:
\begin{itemize} 
\item a general introduction to laboratory work in physics, and within the undergraduate degree scheme at Stockholm University, emphasising the nature of experimentation;
\item the laboratory safety sheets; 
\item an introduction to keeping laboratory notebooks;
\item a very brief introduction to uncertainties and basic uncertainty propagation, based on the pragmatic introduction by Hughes and Hase \cite{Hughes-Hase};
\item a summary of the first year laboratory work, highlighting the special nature of the thermodynamics lab.
\end{itemize}
This background material preceded the laboratory manuals for the entire of first year in a folder (ring binder) that served the purpose of a sturdy, although not rigidly bound, `laboratory notebook'. With the exception of the transformed thermodynamics experiment, the material for the first year experiments was as had been used in previous years. Having developed the material to support the transformed first year labs, it was noted that this material could be useful for students through their degree.

\subsection{The specific heat capacity experiment}

For what was essentially a new experiment, new experimental documentation was needed. This included the laboratory manual and any related activities. Here we describe the elements introduced and the major considerations that led to the implementation realised. First, though, we provide a brief description of the specific heat capacity experiment: since it is a current experiment, we do not include example data or details of known nuances.

The original goal of the experiment and, the pre-determined task of the first session, is to obtain a value for the specific heat capacity of a sample. Students choose a small or large cuboid block of either aluminium, brass or copper. A resistor (one of two sizes) is attached to one side of the sample and acts as a heat source; uncalibrated type K thermocouples attached to each side of the sample provide voltage (mV) readings that are then converted to temperature changes by assuming a constant Seebeck coefficient. Taking readings during heating and cooling, with and without the sample, provides data from which the specific heat capacity of the sample can be calculated.

\subsubsection{Experimental documentation}

We could have had a traditional arrangement with a `script' or `manual' containing the theoretical background (including derivation) and a description of the experiment, from which the students create their introduction and method. However, in multi-session experiments, students can spend the entire of the first session reading, distilling and even copying out large sections of the script into their laboratory notebooks, which is a complete waste of time \cite{PhysEd.53.015016}. Moreover, with only two laboratory sessions in the thermodynamics course, and a total of seven in the first year, laboratory time was precious.

After some debate, and considering the example of scaffolded, pro-forma laboratory notebooks used at Imperial College London \cite{Imp-visit}, in which students are given a document with blank pages and appropriate headers, we decided to create a combined laboratory manual/notebook, hereafter referred to as the `logbook'. We took this idea one step further and, by first creating an example laboratory notebook for the introductory parts of the experiment, created a pro-forma laboratory notebook for the students. This included all the relevant introductory material and derivations, after which the notebook consisted mostly of headers, empty tables (for the first session) and blank pages with some explanations and reminders, as at Imperial.

As mentioned already, the relevant derivation is somewhat long and caused problems in previous years when it was part of the preparatory activities, but it is necessary for the complete documentation of the experiment since it indicates the measurements that need to be made. The full derivation and the formulae for propagating the relevant experimental uncertainties were therefore included in the introductory section of the logbook. The two session experiment meant that there were essentially three parts to the logbooks: the printed introductory section; the structured pages for the first laboratory session; the structured pages for the second laboratory session.

\subsubsection{Preparatory activities}

In order to maximise the time students spent doing the experimental work, it was imperative that students came to the laboratory sessions ready to start doing the experiment. Students can benefit from a chance to practice specific elements of the basic data analysis or key interpretative tasks beforehand \cite{PhysEd.53.015016}, in this case, students practiced the first stage of the data analysis, converting mV readings to temperature changes. 

Having provided the derivations that previously formed part of the preparatory work, and wanting to emphasise the nature of experimentation and promote independence, it was decided that creating a basic outline of the experimental procedure could form part of their preparation for the first session. As detailed earlier (Table \ref{tab:PriorExp}), many students had experience of developing their own experimental methods so this was felt to be a reasonable activity. 

As well as the practice calculation and creating a preliminary method or experimental strategy (ideally somewhat more detailed that the description provided earlier), a third, slightly different element, completed the preparatory work for the first session: completion of the `equipment list' with the purpose of each piece of equipment. This activity relates to the more minor consideration that a long-term outcome would be an improved understanding of instrumentation issues. 

All three elements of the preparatory work relate directly to doing the experiment. These preparatory activities were incorporated into the logbooks and completion confirmed at the start of the laboratory session by a whole group discussion that introduced the experimental equipment and agreed the general experimental strategy. Students contribute the information while the teaching assistant facilitates discussion and checks that no key points are missed.

In the second session, the students' task was to perform their own investigation based on the experiment already performed. Some ideas of directions are given in the documentation, but the only concern for the laboratory assistants in the second session is that the proposed investigation is safe. To find that an experiment does not work is not unusual, and as such, success is not so much that the experiment works as expected, but that the investigation is attempted and valid conclusions drawn. 

The preparatory activity for the second session was, in some respects, simpler than for the first, being for students (working in pairs to do the experiment) to decide upon what they were going to investigate and how they were going to do so. An implicit aspect here is that students complete the analysis and discussion of the first session's work in the time (two and a half to three weeks) between the two laboratory sessions.

\subsubsection{Formative feedback on notebook keeping}

The laboratory part of the thermodynamics course is a pass-fail element requiring the completion of set tasks. This may at first seem unpromising for assessing quality work, but this structure has a great advantage for both skill development and inquiry work. There is no penalty for getting an `incorrect' answer or an experiment not working, thus potentially reducing students' fear of a negative effect on their `performance' if they deviate from set tasks. This format also opens up the possibility for self-guided review and development without the need for any teaching hours to check for honesty since, for example, admitting to `poor quality' note taking (recording of the experiment) carries no penalty.

The conceptual background for the formative feedback tasks was two ideas from UCL: `Traffic Light Feedback' (TLF) \cite{Daven-TLF} whereby the \textit{quality} of a student's work is assessed on a formative scale of absent, inconsistently present and consistently present and the `Data Retrieval Test' (DRT) where students are asked to find relevant pieces of information in their laboratory notebooks \cite{BioSciEd.16.1,ArXiv:1905.13006}.

We created a short online review activity consisting of `survey' questions where any answer was marked as correct, the feedback from which could be used by students to identify what they were doing well and decide what areas of their laboratory notebook keeping they wished to work on. Completion of the activity is a required element of the pass-fail nature of the laboratory work, but there was no penalty for missing information or poor habits.

The activity consisted of two parts: the first is DRT-like and asks for pieces of information such as the mass of the sample, dates, relevant time intervals, or, in the second session, a short summary of the experimental work and its conclusions. A TLF section then addresses quality aspects. In the first session, the quality aspects are about habits of recording, those related to the second session also include a review of new tables. An example of a TLF question and its feedback is given in Fig. \ref{fig:TLF-eg}

{\begin{figure}[htb!]
\center{\includegraphics[width=1\columnwidth]{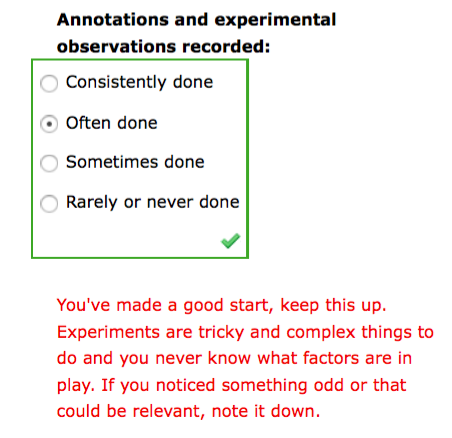}}
\caption{An example of one of the Traffic Light Feedback questions used for both sessions. The four level rating was as a result of comments made during the testing stage. \label{fig:TLF-eg}}
\end{figure}}

The feedback statements provided were always positive and encouraging, with notes explaining why specific points were important and, where changes perhaps may not be obvious, suggesting a strategy that students may wish to try. For example, if students had found it hard to find what might be considered key information, such as the mass or material of the sample used, the feedback suggested to try using stars, boxes or colours to highlight information. 

As well as performing the online part, students also fill in a section at the end of the laboratory session's part of the logbook with two things that they have been doing well and two (possibly small) aspects that they wish to focus on improving. This provides focus for their development and a record that is available in the subsequent laboratory session. This record also provides a means for the teaching assistants to check that the students have completed the activity.

This online, automated formative feedback on laboratory notebook keeping is particularly straightforward in pass-fail laboratories, such as the situation here, where completion only is required, but its use in graded laboratories should be straightforward.

\subsubsection{Overview of the transformed laboratory}

The new laboratory work associated with the thermodynamics course is completed over the course of several weeks, as determined by the scheduled laboratory sessions. The individual elements are described above, and the activities are organised as follows:
\begin{itemize}
\item \textbf{Preparation for the first laboratory session:} read the introductory material; practice with basic data handling; complete the equipment list with purpose of the equipment; create an outline of the basic experimental method (strategy).
\item \textbf{First laboratory session:} review preparatory activities; develop full method for basic experiment; set up equipment; take preliminary data (partial experiment) and discuss; full experimental run; take the experiment apart.
\item \textbf{After first laboratory session:} complete data analysis and draw conclusions.
\item \textbf{Review first laboratory session:} online combined DRT and TLF activity; identify areas to focus on.
\item \textbf{Preparation for the second laboratory session:} discuss and decide on independent investigation; plan experiment; create method.
\item \textbf{Second laboratory session:} check method safe; set up equipment; collect data; take experiment apart.
\item \textbf{After second laboratory session:} complete data analysis, review results and complete experimental record.
\item \textbf{Review of second laboratory session:} online combined DRT and TLF activity, including summary of independent investigation.
\end{itemize}

\subsection{Testing}

As recommended by our visit to Imperial College London \cite{Imp-visit}, after developing the bulk of the material for the transformed laboratory work, we recruited a pair of second year students who had taken the thermodynamics course the previous year and asked them to complete the new laboratory. This was an extremely valuable exercise. It confirmed that the material created to support experimental work overall was appropriate and accessible, and found several weaknesses in the documentation of the experiment. Moreover, it provided valuable experience for the teaching assistants who would deliver the laboratory course, and refined the organisation of the first session in particular to something slightly more rigidly organised.

Specific points that came up during the testing were:
\begin{itemize}
\item incorrect order of magnitude in data for practice data analysis calculations;
\item the need for a well structured first session -- this then provided an ideal opportunity for discussion of how to estimate uncertainties without emphasising their use in calculations, as well as reflecting how experiments often use preliminary data;
\item a clear idea of how the laboratory sessions would run: particularly useful for creating a delivery timetable that was accurate even in the first lab session;
\item how to describe the combined laboratory manual and notebook that had been developed;
\item to have several different coloured folders (ring binders) with the laboratory materials. This would reduce the likelihood of students taking the wrong folder and make the laboratory more inviting by adding a bit of colour;
\item that the three level TLF `rarely or never done; inconsistently done; consistently done' was not nuanced enough, leading to the replacement of `inconsistently done' with `sometimes done' and `often done', as in the example in Fig. \ref{fig:TLF-eg}.
\end{itemize}
The testing stage also confirmed that the laboratory was indeed an improvement on the previous version. We also took the opportunity to ask the students testing the new laboratory work what they felt could be done with the laboratory work for the other courses.

This preliminary implementation stage allowed a very \textit{rapid} evaluation and analysis to adjust the development in a timely manner. It also provided an opportunity for the teaching assistants to practice delivering the new laboratory. The two major differences between the trial run and the final version were that less emphasis was placed on the full record keeping, and the interval between the experimental sessions was only a week. However, any interval of time sufficient to allow the students to perform a rough analysis and review should be sufficient for testing.

\subsection{Preparing Teaching Assistants}

The inquiry laboratories are very different to what the majority of teaching assistants will have experienced in their own studies. Teaching assistants therefore need to prepare carefully to ensure that they can deliver them effectively and appropriately \cite{PRSTPER.10.010116, IntJScholTandL.3.2.136}. This requires attention to several areas:
\begin{itemize}
\item understanding of the purpose of the laboratory for the students' development and how this contrasts with their own experience;
\item prepared delivery: structure for the first session;
\item asking questions to help students answer their own;
\item awareness of the nuances of the experiment, but not pointing them out if students don't find them;
\item agreed standards for aspects such as laboratory notebook keeping and data analysis approaches.
\end{itemize}
These aspects all contribute to the teaching assistants' confidence in delivering the laboratory work in a way that will maximise students' success both in performing experimental tasks and in gaining skills and experience. The coaching and training of teaching assistants should occur in preparation for the experiment. 

If there are significant changes requiring testing, then the testing event provides the ideal opportunity for the teaching assistants to gain experience and understanding of the laboratory in a less demanding and less pressured environment than when delivering it `live' to students. Communication and explanation of the laboratory work, it nature and purpose, to both students and teaching assistants is important. Students should understand why they are being asked to do things; teaching assistants should need to be able to provide effective and appropriate support. 

\section{Reflections on the initial run}
The first run of the new laboratory has been encouraging with students engaging with the experimental work and identifying at least one of the known nuances of the experiment even in the initial data collection phase. While some refinement of the delivery will inevitably occur, the transformed laboratory work seems to be working well and the current intention is to proceed with developing the laboratory work for the other first year modules along similar lines, and using most of the method described here.

\begin{table*}[htb!]
\begin{tabular}{@{} l | p{0.25\textwidth} p{0.25\textwidth} p{0.25\textwidth} @{}} 
\textbf{Aspect} & \textbf{Old experiment} & \textbf{New first session} & \textbf{New second session} \\ \hline
\textbf{Task/Question} & Provided & Provided & Students decide \\
\textbf{Theory/Background} & Provided -- lecture content & Provided in documentation & No additional theory provided \\
\textbf{Experimental setup} & Provided -- already set up & Provided -- students choose some components and set up & Basic set up from first session; students can modify \\
\textbf{Procedures} & Provided -- instructions & Students construct and discuss with teaching assistants & Students decide and confirm safety with teaching assistants \\
\textbf{Results analysis} & Provided in session -- \newline corrected preparatory activity & Basic formulae provided; \newline students decide how to apply & Basic formulae provided, but no constraint to use \\
\textbf{Results interpretation} & Provided -- \newline compare to standard values & Not provided; some prompting to understand raw data & Not provided \\
\textbf{Conclusions} & Guided & Not provided & Not provided \\ \hline
\textbf{Inquiry Level} & \textbf{Confirmation -- Structured} & \textbf{(Guided --) Open} &\textbf{Open -- Authentic}\\ \hline
\end{tabular}
\caption{Characterising the level of inquiry in the old and new thermodynamics experiment on the specific heat capacity using the rubric of Buck et al. \cite{JColSciTeach.38.52}. \label{tab:inq-comparison}}
\end{table*}

In reviewing the work done, we can use the rubric of Buck et al. \cite{JColSciTeach.38.52} to determine the inquiry level of both the old and new laboratory sessions, constructing Table \ref{tab:inq-comparison}. The level of inquiry that students achieve in the second session will vary. Since the samples used for the old thermal conductivity experiment are available for use in the second session, students can investigate an entirely different aspect if they so wish. Note that a lot of the inquiry aspect is achieved quite simply by letting students make decisions about how they do the experiment, giving them permission to modify it, and \textit{not} drawing attention to nuances that are not captured in the basic theory. 

Many students choose to simply modify their method in the second session. This is not to be discouraged, particularly since changing the sample size or geometry while seemingly a small change leads to a real opportunity for students reflect. However, in most laboratory groups at least one pair chooses a less obvious investigation, that can require some reconfiguration of the experimental set up. Examples include measuring the specific heat capacity of a pencil lead (graphite), and investigating the effect of the thermally conducting grease that is used between the resistor and the sample. The implications of these experiences for laboratory work in later years, and indeed later in first year, remains to be seen.

One particular feature that has been seen in student accounts of their investigations is that the second session has not always gone exactly to plan, or yielded the expected results. The structure implemented here seems to be appropriate for letting students fail -- for things to go wrong -- without adding additional pressure to students. The `labs in course' structure which means that the laboratory work is a pass-fail component on completion may be a particular aspect of allowing failure. However, the two session format also allows success: if the first session does not go well, students have the time and opportunity to learn from their experience and succeed in the second session.

Student feedback from the module evaluation indicated that students generally liked the new laboratory format, particularly the freedom of the second session. Opinions were split about the logbook: of the 20 students that completed the survey (50 who completed the laboratory work), half thought the logbook was good, while the other half held diametrically opposite opinions. Overall, the comments indicate that good communication with respect to the goals and format of the laboratory work as part of their training are needed to enable all students to make the most of the open-ended nature of the labs. Changes will be made to the documentation, and the second session will have higher expectations for students to actively test an idea related to the experimental context, thus being more than a simple repeat of the first session's experiment.

While we are generally satisfied with the new experimental work, some concerns remain. There are only eight teaching hours dedicated to the laboratory work for the module, and no additional support or lectures beyond the summaries provided in the general introductory material (running to a total of 27 pages of content including the laboratory safety sheets). This structure makes placing the emphasis on student control and decision making, including about data analysis, which may often be completed at home, easy. However, only the time physically in the laboratory is supported by the teaching assistants. Thus the data analysis methods used are relatively unknown and almost completely unguided. Without knowing what data the students use for their calculations, teaching assistant support within the laboratory is limited since there is a lack of knowledge about what are the most useful or relevant measurements for the students.

This highlights one of the key difficulties that can occur when laboratory work is relegated to a supplementary position and the need for additional support. Since the emphasis of these laboratories is about exploration, and the relevant calculations for analysis are provided, these concerns are, relative to the overall aims of the particular experiment realised, relatively minor. They become vital when the experimental structure described here is used in a more demanding laboratory course. It is not a matter of telling students that one particular approach to data analysis is correct (or better) than another, but allowing them to choose one approach (and be able to explain their choice) and then facilitating a discussion of different possibilities.

\section{Summary and conclusions}

We have recounted in detail the process by which a confirmation laboratory activity was transformed into a genuine experiment allowing students to perform authentic inquiry \cite{JColSciTeach.38.52}. This process may seem daunting, but, when done in a structured manner and with careful discussion and agreement of the aims, is not as difficult as it may at first seem. Moreover, such a transformation can be achieved with very little investment in new equipment, since many existing demonstration `experiments' may be amenable to open-ended student inquiry.

The key conceptual point that facilitated the transformation was that of `Secret Objectives' \cite{ArXiv:1905.07267}, or the identification of experimental nuances that students are \textit{not} told about. Scaffolded laboratory notebooks (logbooks), containing the theoretical description relevant to the initial task provided structure for the students and ensured that preparatory activities were preparation for doing the experiment. In particular, when the method can be inferred from the theoretical description or information about the set up, students can create it themselves. They are put in charge of developing their own experiment from the start.

The new laboratory work has two aims: to introduce students to genuine experimentation, and to provide some training in logbook keeping. Key points in the design of the structure were to ensure success factors while promoting independence by first providing a set experiment where students made decisions about the method and the details of the set up (e.g. sample, resistor type), before they have the option to investigate any aspect they wish to. All parts of the laboratory activities are directly linked to one of the two aims. This structure allows students to gain confidence while also permitting investigation in any direction, thus potentially circumventing the resistance of students to the introduction of inquiry laboratory work \cite{IntJScholTandL.3.2.136}.

While we recognise that every situation is different, we believe that the process we followed is generally relevant and the structure of the new thermodynamics laboratory sessions may be directly applicable to other situations. Here a certain amount of control of the structure of the experimental work was kept, although this may not always be required, and laboratory activities without a laboratory manual can be successful \cite{EJP.39.025703}

\begin{acknowledgments}
We are indebted to P. A. Bartlett, D. A. Armoogum, N. Nicolaou, D. Thomas and A. Korn at UCL, S. Bland, R. J. Forsyth, S. P. D. Mangles, R. A. Smith, G. C. Axtell, H. H. Dawda, and R. E. Whisker at Imperial College London for taking the time to show us round their teaching laboratories and discuss their structure, issues and development in detail. We are particularly grateful to P. A. Bartlett and S. Bland for organising our visits, and sharing examples of their experimental documentation. We would also like to thank Iman and Gustav, the two second year students who tested the transformed laboratory task, and provided the teaching assistants with a valuable (or invaluable) trial run.

The work was supported by a Stockholm Rectors' award for quality development of teaching -- round 5, 2019 (SU FV-1.1.9-0314-15). The visit to UCL and Imperial College London was funded via Erasmus+ funding for Staff Mobility for Training.
\end{acknowledgments}

\bibliographystyle{unsrt}
\bibliography{teachingbibliography_v2}

\end{document}